\documentclass[referee]{raa}            

\usepackage{graphicx,times}  
\usepackage{graphicx}            
\usepackage{natbib}
\usepackage{amssymb,amsmath}
\bibpunct{(}{)}{;}{a}{}{,}
\newcommand{\speed}[1]{#1 km~s${}^{-1}$}

\usepackage[pagebackref=true]{hyperref}

\begin{document}

   \title{First Time Observed M-Shaped Coronal Mass Ejection Associated with a Blowout Jet and an Extreme Ultraviolet Wave}

   \volnopage{Vol.0 (20xx) No.0, 000--000}      
   \setcounter{page}{1}          

   \author{Yu-Hu Miao 
      \inst{1}
   \and Lin-Hua Deng
   \inst{2}
   \and Chao-Wei Jiang
      \inst{3}
   \and Abouazza Elmhamdi
      \inst{4}
   \and Jiang-Tao Su
       \inst{5}
   \and Ming-Xiang Guan
      \inst{1}
   \and Hai-Xin Zou
      \inst{1}
   \and Jiao-Man Li
      \inst{6}
   \and Xue-Mei Cao
      \inst{1}
   \and Jun-Tao Wang
      \inst{7}
    \and Yun-Zhi Hua
      \inst{1}
   }
 \institute{School of Information and Communication, Shenzhen University of Information Technology, Shenzhen 518172, China; {\it miaoyh@sziit.edu.cn}\\
        \and
             School of Mathematics and Computer Science, Yunnan Minzu University, Kunming 650504, China; {\it linhua.deng@ymu.edu.cn}\\
        \and
             State Key Laboratory of Solar Activity and Space Weather, School of Aerospace Science, Harbin Institute of Technology, Shenzhen 518055, China\\
        \and
             Department of Physics and Astronomy, King Saud University, PO Box 2455, Riyadh 11451, Saudi Arabia\\
        \and
             School of Astronomy and Space Sciences, University of Chinese Academy of Sciences, 19 A Yuquan Road, Shijingshan District, Beijing 100049, China\\
        \and
             School of Digital Media, Shenzhen University of Information Technology, Shenzhen 518172, China\\
        \and
             School of Physical Science and Technology, Lingnan Normal University, Zhanjiang 524048, China\\
\vs\no
   {\small Received 2025 Aug 23; accepted 2025 Oct 27}}

\abstract{ The coronal blowout jet, extreme ultraviolet (EUV) wave and coronal mass ejection (CME) are common phenomena in the solar atmosphere.
In this paper, we report the occurrence of an M-shaped CME event associated with a blowout jet and an EUV wave using high-resolution,
multi-angle and multi-wavelength observations taken from {\sl Solar Dynamics Observatory}, and {\sl Solar TErrestrial RElations Observatory}. Interestingly, and for the first time, it is found that two bubble-like CMEs and a jet-like CME were simultaneously triggered by the same eruptive event. Our observational analyses and findings indicate the following: (1) the eruption of a blowout jet led to a large-scale EUV wave; (2) the eruption of the EUV wave swept a small filament (prominence) and a long filament; (3) eventually the EUV wave split-up into two parts, leading to the two bubble-like CMEs, while the blowout jet induced a jet-like CME. The combined events appear to form an M-shape like structure CME, that we sketch throughout a proposed cartoon tentatively explaining the observed complex configuration. Based on observational diagnosis, we argue that the jet, the EUV wave and the multi-CME are highly interlinked. A suggested eruption-model, from the solar atmosphere to the space, is outlined and discussed, providing a possibly new way to probe the relationship between the solar eruptions and the surrounding space. 
The investigation of such rare phenomenon can be a key point for better understanding of the physical associated triggering mechanisms and energy transport in the solar atmosphere, crucial for MHD simulations and modeling.
}
   \authorrunning{Miao et al.}            
   \titlerunning{ }  
   \maketitle

%
%
\section{Introduction}           
\label{sect:intro}

The short-lived coronal jets are frequent phenomena in the solar atmosphere \citep{shen2017,shenyd2018b,shenyd2021,yanglh2023}. A primary standard two-dimensional jet model was first presented by \citet{shib92}. In the recent two decades this model has been widely studied in two-dimensional and three-dimensional simulations and subsequently adopted by the scientific community \citep{yoko95,yoko96,liu04,li17,shenyuandeng2019}. Magnetic reconnection is known to play a crucial role in the acceleration of solar jets \citep{shib94,shen11,shenyd2018,duanyd2019,sterling2020,chenhc2021,shenyd2021,wangya2021,zhoucr2021}. \citet{liu04} analyzed a surge that was driven by emerging magnetic fluxes associated with low reconnection on the photosphere. Interestingly and about a decade ago, \citet{moor10} found a new type of coronal jet that was dubbed ``blowout jet''. \citet{zhu17} reported that a blowout jet can be triggered by the eruption of a magnetic flux rope.

Various investigating efforts have shown that coronal jets could result in CMEs \citep{liu05,liu08,wang98,miao2018,nova2019,miao2019a,miao2019b,Solanki2020,tianhui2021}. \cite{panesar2016} reported that jets can trigger CMEs to erupt. \cite{liu11} studied the on-disk observations of a blowout jet exhibiting an intriguing loop-blob structure. The authors argued that the erupting blob and loop might be a miniature version of CMEs in the low corona as proposed earlier by \cite{moor10}. \citet{shen12}, and \citet{miao2018} analyzed a blowout jet eruption associated with a rare double-CME event, respectively. The authors presented a jet-like CME and a bubble-like CME evolving from a single blowout jet. They suggested that the jet-like CME was directly produced by the hot part of the blowout jet whilst the bubble-like CME was caused by the mini-filament eruption or the reconstruction of the high coronal fields. The jet-like CME and the bubble-like CME were in side by side \citep{shen12} or in overlapping \citep{miao2018} relationship along the line of sight from the FOV of {\sl STEREO}. More recently, \citet{tang2021} presented an interesting event namely a standard jet and a blowout sympathetic jet originated from two adjacent coronal bright points.

Worth to mention here the relevant discovery, namely the detection of the global propagating EUV waves, first time revealed by the Extreme-ultraviolet Imaging Telescope (EIT) onboard the {\sl Solar and Heliospheric Observatory spacecraft} \citep[{\sl SOHO};][]{dela95,moses97,thompson98}.
Most of the recent well observed and inspected EUV waves were accompanied by large-scale energetic flares, and they were thought to be fast-mode magnetic waves driven by CMEs \citep{chenpf2002,chenpf2002a,shenyd2012a,shenyd2012b,shenyd2013,shen2014a,li12,yang13,chenpf2016,wangcan2021}.
Over the past few years, solar physicists have been debating about the wave and non-wave nature of the observed EUV waves,
as well as on the possible associated triggering mechanisms induced by flares or/and CMEs events \citep{lu2021,zhouxp2021,miao2025}. \citet{zhangjie2004} reported three CME events where two events associated with flares and one event without any flare. indeed, these critical issues are yet still subject of current hot debates and ongoing research inspections. Exploiting the interaction between EUV waves and other coronal structures can be very important to further scrutinize the physical
nature of the EUV waves \citep{shen2014a,shen2014b,miao2020,miao2021a,radoslav2022}. The EUV waves have been observed to interact with a variety of coronal structures \citep{thompson98,thompson99,ofman02,veronig06,wangya2018,miao2021b}. \citet{gop09} presented the first example of reflection and refraction of EUV waves at a remote coronal hole. \citet{patsourakos2009b} proposed a model to interpret the relationship of the EUV and the CME.

In this paper, we present the first observational evidences of an M-shaped CME event occurred on March 10, 2011.
This very rare event appeared at the southeast of active region NOAA AR11166 near disk center as viewed from the {\em SDO} and
{\em STEREO} Ahead (STA) and Behind (STB). In the following, we particularly focus on highlighting and exploring the M-shaped CME event, whereas the EUV-wave detailed analysis was presented in \citet{miao2019b}. In this paper, we report an M-shaped CME event, which have a close relationship with a blowout jet and EUV waves. Instruments and observations are described in Section 2. Analysis and results are depicted in Section 3, while discussions and conclusions are summarized in Section 4.

\section{Observations}
\label{sect:Obs}

The observations used in this paper are provided by the {\em SDO}/Atmospheric Imaging Assembly (AIA; \citealt{lemen12}), the
{\sl STEREO}/Extreme Ultraviolet Imaging Telescope (EUVI; \citealt{wuelser2004}), and the COR1 coronagraph \citep{thomp03}.
On 2011 March 10 at 06:30:00 UT, the separation angle between STA and STB was about $177^\circ$, while that between STA (STB)
and {\em SDO} was about $88^\circ$ ($95^\circ$). Since the event occurred near the disk center in the {\em SDO} field-of-view (FOV), it was located on the east and west disk limb in the FOVs of STA and STB, respectively. The AIA onboard the {\sl SDO} has 12-second cadence and exposures of 0.12-2 seconds, and it images the Sun up to 1.3 R$\sun$ in seven EUV wavelength bands with a pixel width of $0\arcsec.6$. Here, we primarily use the 171, 193, and 304 \AA\ AIA channels. In addition, we benefit of the surface magnetic field data supplied by the {\sl SDO}/Helioseismic
and Magnetic Imager (HMI; \citealt{sche12}), especially, the Space-weather HMI Active Region Patches (SHARPs; \citealt{bobra2014}) for the NLFFF extrapolations purpose. The EUVI observations from the Sun Earth Connection Coronal and Heliospheric Investigation (SECCHI; \citealt{howard08})
onboard {\em STEREO} are also used. It takes full-disk 195 \AA\ and 304 \AA\ images, which have a 5 and 15 minute cadence and a pixel width of $1\arcsec.6$, respectively. The COR1 coronagraph observes the corona in 1.5-4 R$\sun$ with a 5 minute cadence and a pixel resolution of $15\arcsec$.

\begin{figure}
 \centering
\includegraphics[width=0.8\textwidth]{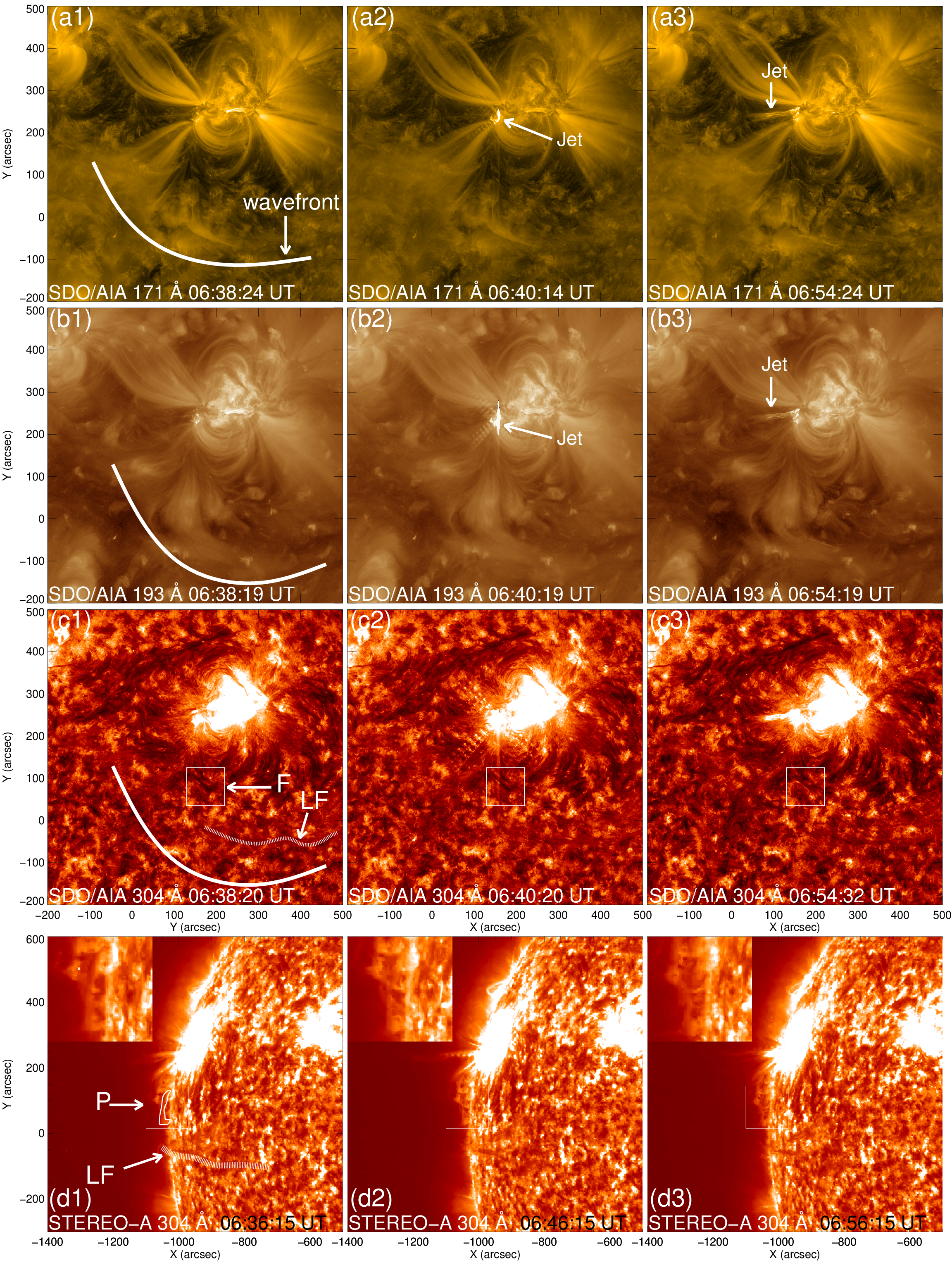}
\caption{The white curved lines indicate the EUV wave propagation path and profile at the AIA 171, 193 and 304 \AA\ channels. The blowout jet is also shown in panels (a2), (a3), (b2) and (b3). The white boxes indicate the small filament (prominence) evolution in panels (d1)-(d3). The bottom three panels refer to the observations from the FOV of STA at 304 \AA. The white boxes mark the small prominence (filament) designated with the ``P'' symbols. The white contour line represents the profile of the small prominence. ``LF'' represents the long filament structure that indicated by the white arrow. The zoomed views highlight the prominence evolution. An accompanying movie (animation1.mpeg) is available in the online journal.
\label{euv}}
\end{figure}

In order to highlight the evolution of the jet and the EUV wave, we report of Figure \ref{euv} two views, one from the {\em SDO}/AIA data, and one from the {\em STEREO}/EUVI data. The upper three rows of Figure \ref{euv} show the {\em SDO}/AIA data. We use white curved lines to represent the evolution of the original EUV wave. The white contour line highlights the profile of the small filament (prominence). The EUV wave is found to sweep up both the small filament(prominence) and the long filament during the time interval 06:41 to 06:45 UT. The bottom row in Figure \ref{euv} shows the small prominence (filament) and the long filament from STA perspective. In Figure \ref{rundiff}, the {\em SDO}/AIA 171 and 193 \AA\ running-difference images report the evolution of the large scale EUV wave. The wave is divided into three parts, one part is the reflected wave, and the other two parts are wave1 and wave2, respectively. To emphasize more the relationship between the jet and CMEs, a set of original EUVI 304 \AA\ and running-difference COR1 combined images from two views of ST are reported in Figure \ref{cor304}. Using running-difference EUVI 195 \AA\ and running-difference COR1 combined images, the association between EUV wave and CMEs is also accentuated from two views of ST in Figure \ref{cor195}. It should be noticed that the time of the CME began to appear in the field of view of COR1 was about 06:55 UT. In addition and to better understand the M-shaped CME event, the related dynamics is analyzed and emphasized in Figure \ref{slice}.


Moreover, and to illustrate the configuration and evolution of the magnetic field, NLFFF extrapolations \citep{jiangchaowei2010,jiangchaowei2018} and PFSS \citep{schatten1969,schrijver2003} modeling are both performed and the results are reported in panel(a), panel(b) and panel(c), respectively. The evolution of the jet is indicated by white arrow in panel (a). The propagating directions of wave1 and wave2 are also marked by black arrows in panel(a). Panel(b) displays only the jet base region magnetic field lines, while panel(c) displays the magnetic field lines from PFSS model overlapped on the photospheric magnetogram. Panel(d) illustrates a proposed cartoon sketching the M-shaped CME configuration from the view of STA.

\section{Data analysis}
\label{sect:analysis}

On 2011 March 10, an M-shaped CME event, which appeared at about 07:55 UT in the STEREO/COR1 field of view and occurred in NOAA active region (AR) 11166. The event is evidently observed to have a close relationship with a blowout jet and EUV waves.
\citet{miao2019b} presented a detailed study of the Quasi-periodic fast-propagating (QFP) magnetosonic wave and the EUV wave event. The relationship of the blowout jet and the M-shaped CME was reported by \citet{miao2019a}, but few details about the CME event were provided.

In Figure \ref{euv}, the white curved lines and the white dotted curve lines indicate the EUV wave propagation profile and
the long filament in 171, 193 and 304 \AA\ channels, respectively. It is interesting that the EUV wave swept up both a small filament (prominence)
and a long filament. Worth mentioning that the small filament (prominence) and the long filament presence and behaviour are commonly considered
as evidences of CME eruptions. Panels (c1)-(c3) of Figure \ref{euv} display the small filament (prominence) evolution from time 06:38:20 to 06:54:32 UT. From the view of the {\sl STEREO} Ahead, a small prominence is clearly distinguishable in the EUVI 304 \AA\ images. Its profile is highlighted by the white contours in the box of panel (d1) of Figure \ref{euv}. Through comparison we confirm that the small prominence, is actually, the small filament from the view of {\sl SDO}. The ``P'' symbol indicates the small prominence in panel (d1) of Figure \ref{euv}. In order to emphasize the evolution of the small prominence, a related zoomed view marked by a box, is displayed in panels (d1)-(d3) of Figure \ref{euv}. It allows one to distinguish the prominence evolution within the studied time interval of interest.
We mention here that the long filament got swept up by the EUV wave at about 06:45 UT (see animation1.mpeg).

\begin{figure}
 \centering
\includegraphics[width=\textwidth, angle=0]{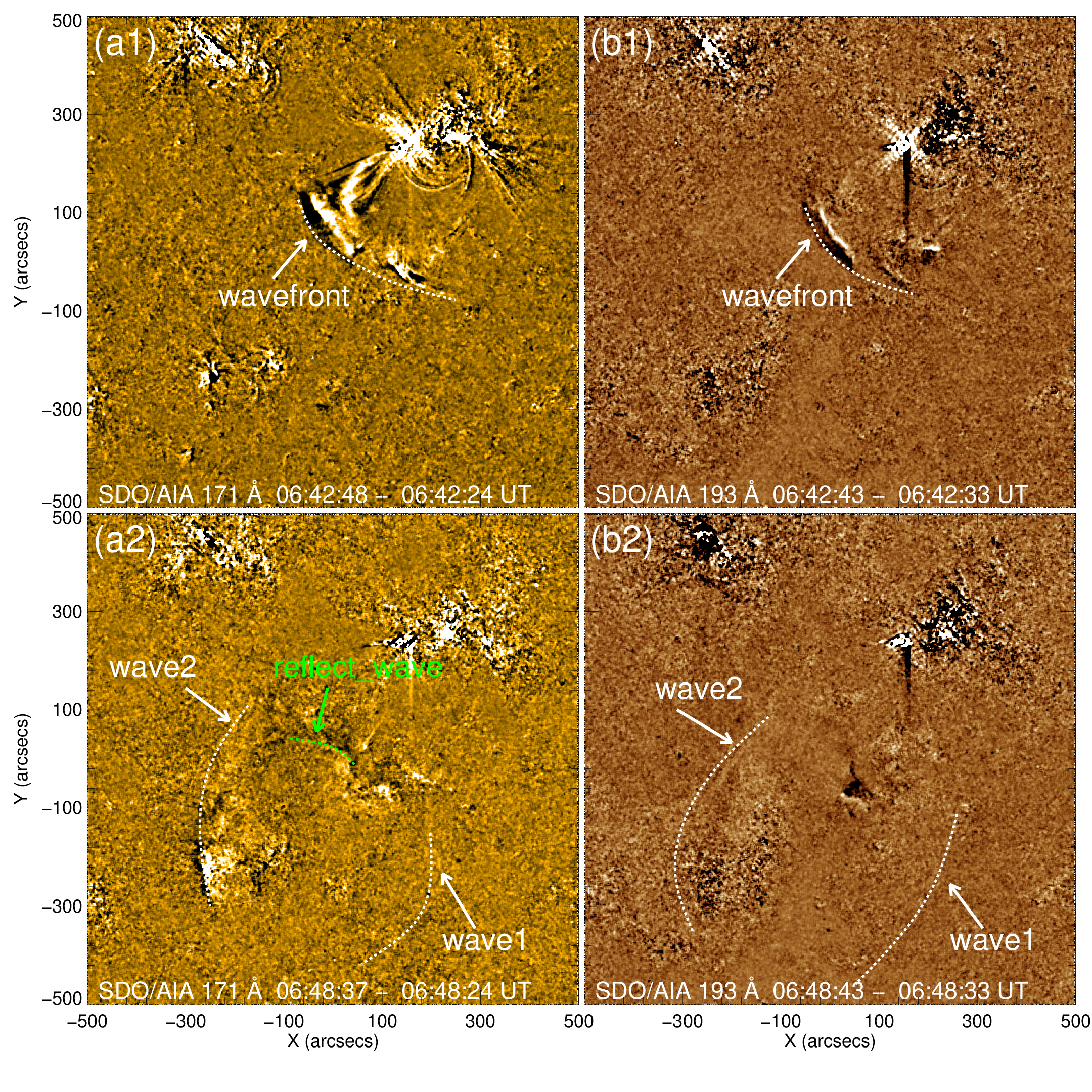}
\caption{The top and bottom panels report the observations from {\em SDO}/AIA 171 and 193 \AA\ running-difference images, respectively. The white-dotted curved lines indicate the wavefront of the EUV wave. The reflect wave is also displayed in panel (a2). An accompanying movie (animation2.mpeg) is available in the online journal.
\label{rundiff}}
\end{figure}

Taking advantage of the {\em SDO}/AIA 171 and 193 \AA\ running-difference images, we present in Figure \ref{rundiff} the initial and the separate
wavefronts of the EUV wave. The EUV wave to sweep up the small filament (prominence) and the long filament from
06:41 UT to 06:45 UT. The evolutions of the EUV wave, the small filament (prominence) and the long filament can be seen in the
online accompanying videos: animation1.mpeg and animation2.mepg.

\begin{figure}
 \centering
\includegraphics[width=4.5 in, scale=01, angle=0]{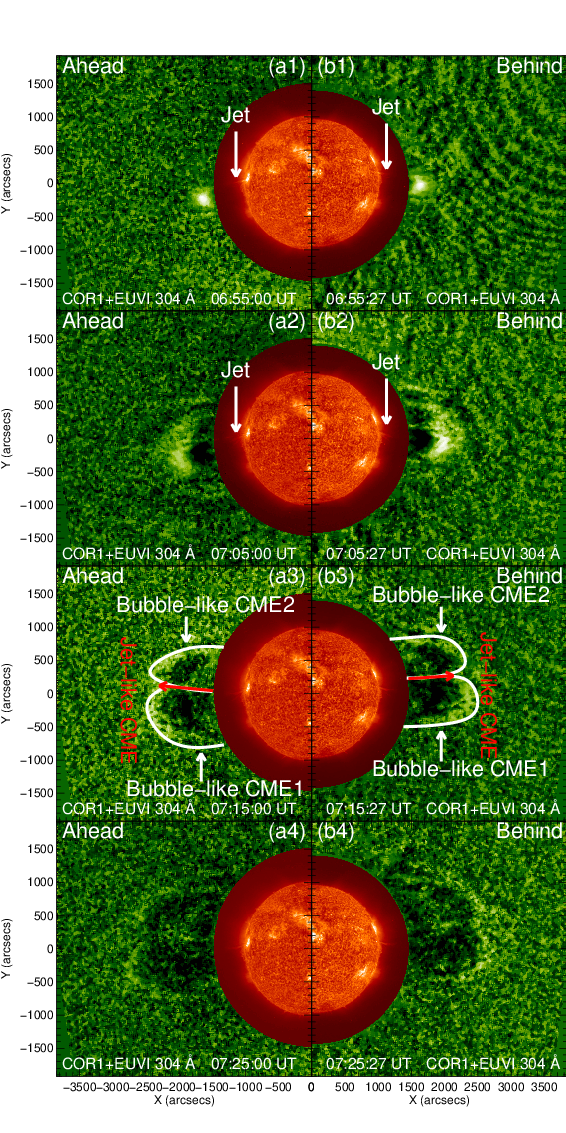}
\caption{A set of original 304 \AA\ and running difference COR1 combined images illustrating the M-shaped CME
evolution from the two viewpoints of STA and STB. The blowout jet is also indicated with white arrows
in the panels (a1), (b1). Panels (a3), (b3) display the structures of the three CMEs. The red arrows indicate
the jet-like CME, while the two white arch lines represent the two bubble-like CMEs. An animated version (animation3.mpeg) of this
figure is available in the online journal.
\label{cor304}}
\end{figure}

In Figure \ref{cor304}, a set of original 304 \AA and running difference COR1 combined views clearly highlight
the M-shaped CME evolution from the two viewpoints of STA and STB, where we indicated the blowout jet in panels (a1) and (b1).
Animation3.mpeg in the online journal contains STA and STB images too. The movie reveals the characterising details of the blowout jet eruption. Very interestingly, the bubble-like (bubble-like CME 2) and the jet-like CMEs were concurrently observed to be associated with the blowout jet in the FOVs of the COR1 STA and STB. Observations by \citet{miao2019b} and \citet{ofman2025} on the same AR 11166 event show that the blowout jet propagates along a funnel-shaped loop system. The jet transitions from compact ejection to diffuse expansion, which is not a sign of energy loss but rather plasma accumulation along the loops. This accumulation process is consistent with the speed increase of the jet in this manuscript-from an initial \speed{158} to a final \speed{511-525}, as the jet accumulates surrounding plasma to form the jet-like CME. Additionally, multi-view simulations by \citet{ofman2025} demonstrate that the same magnetic axis of AR 11166 appears as radial propagation in SDO (disk-center view) and tilted propagation in STEREO (limb view). This simulation result confirms that the apparent direction difference between the jet and the jet-like CME in this manuscript is a projection effect, not a true difference in propagation direction.

\begin{figure}
 \centering
\includegraphics[width=\textwidth, angle=0]{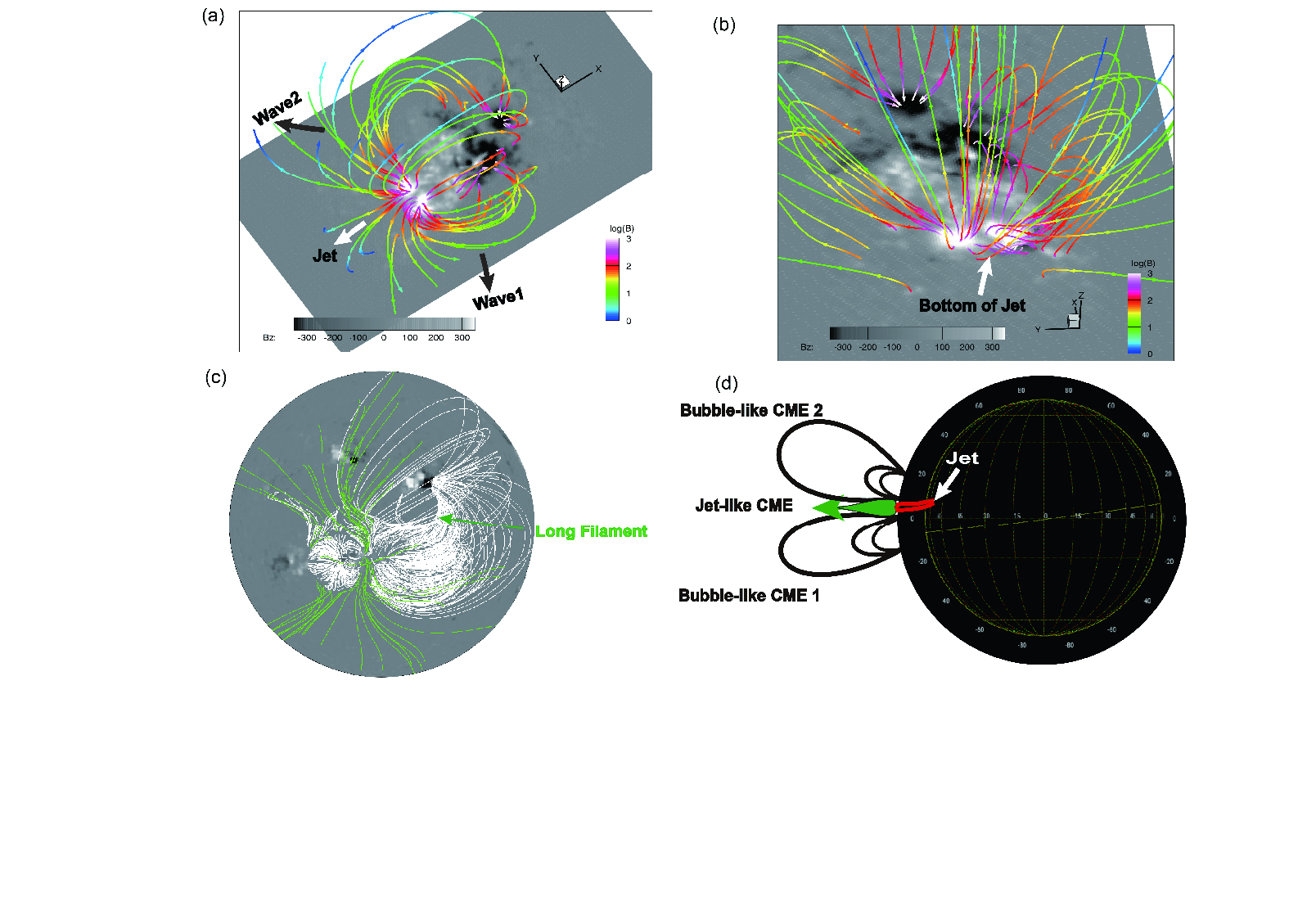}
\caption{Panles (a) and (b) report the results of NLFFF extrapolations. The evolution of the jet is indicated by white arrow in panel (a). The propagating directions of wave1 and wave2 are also indicated by black arrows in panel(a). The panel (b) only shows the jet base region magnetic field lines. Panel (c)
displays the magnetic field lines from PFSS model overlapped on the photospheric magnetogram. The location of the long filament is indicated by the green arrow. Panel (d) illustrates the proposed cartoon (see text) showing the M-shaped CME configuration from the view of STA.
\label{model}}
\end{figure}

This blowout jet eruption associated with simultaneous double-CME event is reminiscent of the simultaneous
double-CME triggered by a blowout jet described and model-interpreted in the work of \citet{shen12}. Around 07:05 UT, the jet-like CME and the bubble-like CME2 could be distinguished in the FOVs of COR1 STA and STB (see panels (a2), (b2)). We highlighted the three CMEs profiles in panels (a3) and (b3) of Figure \ref{cor304}. To confirm the source of CME2, we analyzed AR 11169 via SDO/AIA and STEREO/EUVI observations (06:30-07:30 UT) and found no eruption signatures: no jet, EUV wave, or flare-the bright point in Figure \ref{cor304}(b) is a weak, non-eruptive emission; CME2 first appears in COR1 at about 07:05 UT, $\sim$ 20 minutes after wave2 sweeps the small filament (06:45 UT), matching the wave-driven CME time scale; its propagation direction (white arch in Figure \ref{cor304}(a3)-(b3)) aligns with the direction of wave2 (Figure \ref{model}(a)), and speed (\speed{529-547}) is consistent with the speed of wave2 ($\approx$ \speed{530} from SDO/AIA data), ruling out AR 11169 as a source.

For further validation, observations by \citet{miao2019b} on the same event show no eruptions in AR 11171, a nearby active region of AR 11166. This result suggests non-target active regions similar to AR 11169 do not have the ability to produce eruptions. Additionally, the PFSS model by \citet{ofman2025} reveals that AR 11166 has a closed magnetic loop system. This system prevents propagation of waves and CMEs across active regions, providing magnetic topology evidence that CME2 cannot originate from AR 11169.

\begin{figure}
 \centering
\includegraphics[width=4.5 inch, angle=0]{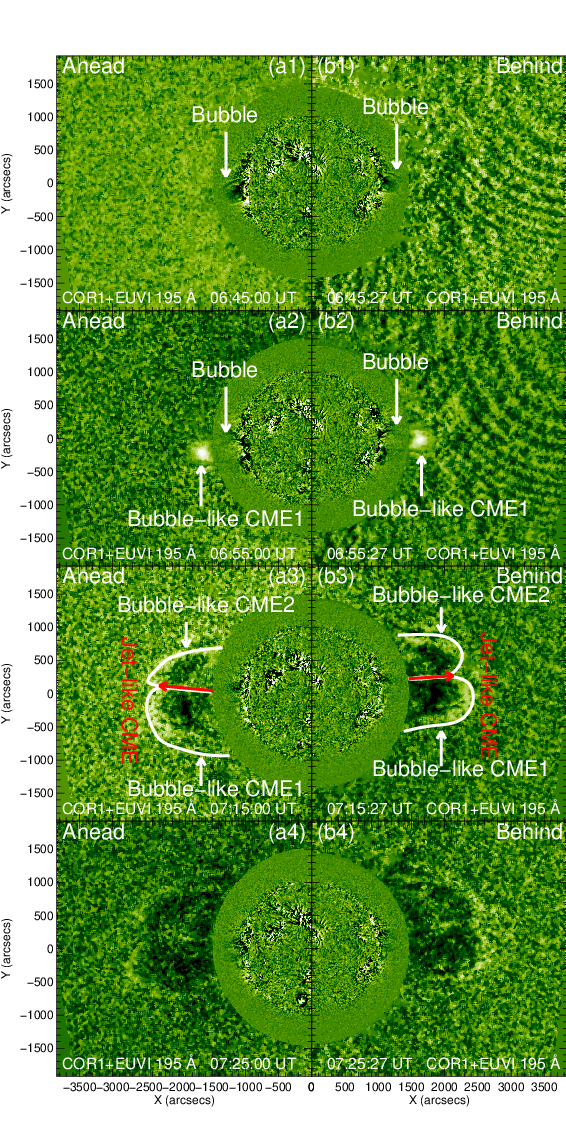}
\caption{A set of running difference 195 \AA\ and COR1 combined images showing the EUVI bubble and the first bubble-like CME (bubble-like CME 1) evolution from the two viewpoints of STA and STB (see animation4.mpeg).
\label{cor195}}
\end{figure}

In Figure \ref{cor195}, a set of running difference 195 \AA, adopting 10 minutes time interval for the image substraction
and COR1 combined images, depict the EUVI bubble evolution from the two viewpoints of STA and STB. Starting at 06:45:00 UT,
the EUVI bubble was distinctly discernible, especially when it emerged from the COR1-A and
COR1-B FOVs at about 06:50:00 UT (see animation4.mpeg). It should be
noted that the first bubble-like CME cavity in COR1 and the EUVI bubble clearly exhibited
the same structure (see all panels of Figure \ref{cor195} from time 06:45 to 07:25 UT). We join an animation
(animation4.mpeg in the accompanying material) to illustrate the detailed characteristics
of the EUVI bubble and the M-shaped CME progressions.

\begin{figure}
 \centering
\includegraphics[width=\textwidth, angle=0]{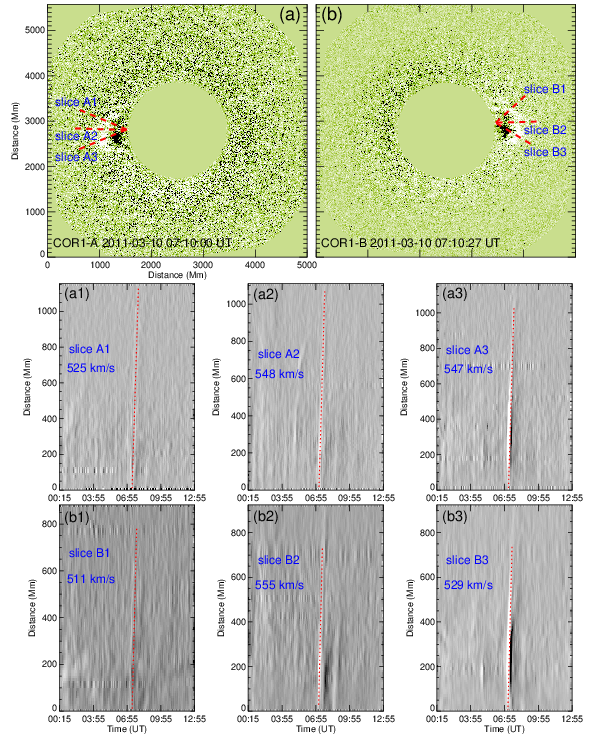}
\caption{STEREO/COR1 running-different images at 07:10 UT from two views in panels (a) and (b). Slices A1-A3 and slices B1-B3 are used to obtain the time-distance diagrams shown in panels (a1)-((a3) and (b1)-(b3), respectively.
\label{slice}}
\end{figure}

As noted in \citet{miao2019a} the blowout jet seemingly triggered the CME event. The authors quantified the speed of the blowout jet to be about \speed{158}, while the speed of the CME was about \speed{263} according to {\em STEREO} CACTus database\footnote{https://secchi.nrl.navy.mil/cactus/index.php?p=SECCHI-A/2011/03/out/CME0034/CME.html \label{cme}}.
Since the speed provided from the CACTus is a rough estimate, we also measured the dynamic parameters of the Triple-Structure CME from two perspectives (see Figure \ref{slice}). The speeds of the jet-like CME and the two bubble-like CMEs are about \speed{511-525}, \speed{548-555}, and \speed{529-547}, respectively. 
To further clarify the splitting mechanism of the EUV wave and its interaction with magnetic structures, observations by \citet{miao2019b} on the same event show that the EUV wave in AR 11166 undergoes local refraction and reflection at magnetic separatrices but that wave1 and wave2 still propagate along the funnel-shaped closed magnetic loops of AR 11166. This result confirms that magnetic topology dominates the propagation direction of the split waves. Additionally, the 3D MHD simulations by \citet{ofman2025} using the potential magnetic field of AR 11166 demonstrate that QFP waves-cospatial with EUV waves-propagate along magnetic loop branches without refraction-induced front bending. This simulation result matches the straight propagation feature of wave1 and wave2 in the SDO/AIA images of this manuscript (Figure \ref{rundiff}), further validating the magnetic topology-guided splitting mechanism.

In order to gain more insights into the associated magnetic field
configuration, we construct and report in Figure \ref{model}, the resulted field extrapolations of a nonlinear force-free field extrapolation (NLFFF) \citep{jiangchaowei2018,zoupeng2020} and a potential-field source-surface (PFSS) \citep{schatten1969,schrijver2003} models. Panels (a) and (b)
show the AR11166 magnetic field lines and the blowout jet base to be highly confined to the closed loops. In the panels (a) and (b), only a few
representative field lines are drawn. Panel (c) of Figure \ref{model} displays the global magnetic topology structure. The low-lying loops and high-lying
loops structures are found to be quite reasonable to interpret the M-shaped CME event, notably, the two bubble-like CMEs. For the purpose to distinctly highlight
the M-shaped CME structures, we present a cartoon in panel (d) of Figure \ref{model}. The blowout jet profile is overlapped in Figure \ref{model}(d).
From the animation, the jet-like CME was directly caused by the blowout jet. In the cartoon, the
green arrow represents the jet-like CME and the two white arch lines refer to the two bubble-like
CMEs, respectively.

\section{Discussions and Conclusions}
\label{sect:discussions and conclusions}
Combining the analysis of high-spatial and high-temporal resolution multi-wavelength observations from three different
viewpoints, we present the so far first M-shaped CME event that was observed on 2011 March 10. Here we mainly
focus on the configuration of the M-shaped CME event. Globally, our primary results in this paper (among various findings)
indicate the M-shaped CME event to be formed by two bubble-like CMEs together with one jet-like CME. \citet{Biesecker2002}
showed that EUV waves emanate from flaring active regions but are strongly associated with the CME. \citet{Patsourakos2009a} studied the relationship of the EUV wave and the CME. The authors indicated that the EUV wave and the CME are originally connected but later-on develop into two separate phenomena. On the one hand, and in our case, however, we consider that the blowout jet triggered the initial EUV wave. Eventually, when the initial EUV wave swept the small filament and the long filament, it splits into two parts. The two bubble-like CMEs were probably caused by the counterpart of the two second waves (wave1 and wave2, see Figure \ref{rundiff}). The first bubble-like CME was stronger than the second one. Because the initial EUV wave swept the small filament and the long filament, near the wave1, most of the material of the small filament and the long filament were injected through the first bubble-like CME.

Observations by \citet{miao2019b} on the same AR 11166 event confirm that the fast component of the EUV wave is a CME-driven piston shock. The kinetic energy of this shock disrupts the magnetic equilibrium of filaments in AR 11166, providing cool filament material as the source for CME formation. Furthermore, 3D MHD simulations by \citet{ofman2025} show that wave propagation in the loops of AR 11166 increases current density j$^2$, which is proportional to Ohmic heating. This heating process raises the temperature of lifted filament material to coronal temperatures ($\sim$ 10$^6$ K), enabling the formation of coherent bubble-like CME structures. These findings from the two studies validate the energy conversion pathway proposed in this manuscript.

On the other hand, the M-shaped CME event itself is very complex and apparently the involved mechanisms are complex too. \citet{shen12}
presented a model to interpret the blowout jet eruption associated with two simultaneous CMEs. The M-shaped CME
case, obviously, is not easy to interpret and explain by adopting that model. Based on our observational results,
the bubble-like CME1 and the bubble-like CME2
were most likely directly caused by the wave1 and wave2, respectively. The jet-like CME was apparently caused by
the blowout jet (see animation3.mpeg). The M-shaped CME event was initially triggered by the eruption of the blowout jet.
\citet{miao2019b} identified the fast EUV component in the same AR 11166 event as a CME-driven piston shock (with speeds of \speed{470-923}), whose kinetic energy disrupts the magnetic equilibrium of filaments in AR 11166 to supply material for CME formation. \citet{ofman2025} further demonstrated through simulations that wave propagation within the loops of AR 11166 enhances current density j$^2$ (a quantity proportional to Ohmic heating). This process heats the lifted filament material to coronal temperatures ($\sim$ 10$^6$ K) and in turn validates the energy conversion pathway proposed in our manuscript.

\citet{chenpf2002} indicated that the fast-component wave is a fast-mode MHD wave and the slow-component one is
possibly formed by successive stretching magnetic field lines. \citet{shenyd2012b} and \citet{miao2019b} also confirmed the coexistence of fast and slow wave components as predicted in numerical studies by \citet{chenpf2002}. Indeed, according to \citet{miao2019b} where the authors studied the same event of the present work, their results indicated that the velocities of the slow-component waves are roughly one-third of the fast-component EUV wave, as predicted by the magnetic field-line stretching model presented by \citet{chenpf2002}. Hence, the EUV wave1 and wave2 were probably formed by the magnetic field-line stretching as illustrated in Figure \ref{model}(a). It is reasonable to interpret the two bubble-like CMEs from the two views of {\em STEREO}. In Figure \ref{model}(b), the one foot point of magnetic field lines of jet, wave1 and wave2 are rooted in the jet base region. Therefore, jet, wave1 and wave2 collectively form the middle part of the ``M'' structure (see Figure \ref{cor304}, Figure \ref{cor195} and Figure \ref{model}(d)). Many studies indicated that magnetic reconnection is the main triggering mechanism of the jet eruption \citep{yoko95,yoko96,liu04,li17,miao2018,shenyuandeng2019}. \citet{miao2019a} demonstrated that blowout jet can possibly trigger the jet-like CME. Above all and combined with Figure \ref{model}, the M-shaped CME is a new kind of CME configuration. It is not only confirming the numerical model by \citet{chenpf2002} but also indicating the magnetic field lines foot point of M-shaped CME all rooted in the jet base region. We regard the M-shaped CME event as a very important finding which can provide new insights into improving the modelling of the jet and the CME phenomena. However, it is wise to keep considering the role/possibility of a combination of complex angles-of views effects.

In our present investigation we principally focus on the morphology of the M-shaped CME event. Our main findings can be summarized as follows.

(1) The initial EUV wave was apparently triggered by the eruption of the blowout jet. The two bubble-like CMEs were
originated from the two second waves (wave1 and wave2, see in Figure \ref{rundiff}), namely, the wave1 led to
the first bubble-like CME (bubble-like CME1) while the wave2 led to the second bubble-like CME (bubble-like CME2).
The jet-like CME was directly caused by the blowout jet.

(2) The blowout jet eruption, leading to a jet-like and two bubble-like CMEs, can be interpreted adopting the scenario presented by \citealt{shen12} and \citealt{miao2018}. The two bubble-like CMEs were most likely triggered according to the mechanism predicted in numerical studies by \citet{chenpf2002}. But the blowout jet associated with M-shaped CME event appears too complex to be explained uniquely through the previous jet available models. Hence, we construct and propose a new model that we dub: "jet-wave-multi-CME``. The hot part of the jet has enough high kinetic energy to drive the EUV wave. Indeed, the EUV wave and the magnetic field-line stretching swept up the cool material and compressed the plasma in the expansion path, such as filaments. The balance of the structures were destroyed by the EUV wave and magnetic field-line stretching, consequently the cool material was heated through the waves escaping the outer corona.

(3) The first M-shaped CME event is an interesting and important finding, and can highly help improving the models of the jets and the CMEs.

Finally, we consider that this first ever reported M-shaped CME event being related to one coronal blowout jet is consisting a very rare phenomenon. This study not only might have significant impact for better future probing and understanding of solar eruptions varieties, but also riches the configuration of CME. More detailed statistical and theoretical research is needed in the future to verify the jet-wave-multi-CME model proposed in the present study.

\begin{acknowledgements}

The authors thank the referee for his/her valuable suggestions that improved the quality of the paper. We thank for the excellent data provided by the {\em SDO} and {\em STEREO}. Y.H.M. is supported by the National Natural Science Foundation of China (NSFC 12103016), the Fund of Shenzhen Institute of Information Technology (Nos. SZIIT2025KJ003 and HX-0951), and High-Talent Research Funding under Grant (RC2024-003). L.H.D. is supported by the National Nature Science Foundation of China (12463009), the Yunnan Fundamental Research Projects (grant No.202301AV070007), the ``Yunnan Revitalization Talent Support Program'' Innovation Team Project (grant No.202405AS350012). M.X.G. is supported by the project of Shenzhen Science and Technology Innovation committee (Nos. KJZD20240903103300002, KCXFZ20240903094011015). A.E. extend his appreciation to the Deanship of Scientific Research at King Saud University for funding this work through research group No. (RG-1440-092). X.M.C is supported by the Fund of Shenzhen Institute of Information Technology (No. SZIIT2025KJ051). J.M.L. is supported by the Fund of Shenzhen Institute of Information Technology (No. SZIIT2025SK035). J.T.W. is supported by the National Natural Science Foundation of China (NSFC 12003005). Y.Z.H is supported by the High-Talent Research Funding under Grant (RC2022-001).
\end{acknowledgements}

\bibliographystyle{raa}
\bibliography{ms2025-0412}

\label{lastpage}

\end{document}